%% file: main.tex
\pdfoutput=1 
\documentclass[journal,transmag]{IEEEtran}

\usepackage{ifpdf}
\usepackage{cite}
\usepackage{amsmath}
\usepackage{algorithmic}
\usepackage{array}
\usepackage{fixltx2e}
\usepackage{stfloats}
\usepackage{bm}
\usepackage{url}
\usepackage[pdftex]{graphicx}

\begin{document}

\title{SIMBA: A \underline{S}kyrmionic \underline{I}n-\underline{M}emory \underline{B}inary Neural Network \underline{A}ccelerator}

\author{\IEEEauthorblockN{Venkata Pavan Kumar Miriyala,~\IEEEmembership{Student Member,~IEEE}, Kale Rahul Vishwanath,
Xuanyao Fong,~\IEEEmembership{Member,~IEEE}}
\IEEEauthorblockA{Department of Electrical and Computer Engineering,
National University of Singapore, Singapore, 117583.}

\thanks{Manuscript received XXXXX; revised August XXXXX. 
Corresponding authors: V. P. K. Miriyala  (email: elevpkm@nus.edu.sg), X. Fong (kelvin.xy.fong@nus.edu.sg).}}

\markboth{Journal of \LaTeX\ Class Files}%
{Shell \MakeLowercase{\textit{et al.}}: Bare Demo of IEEEtran.cls for IEEE Transactions on Magnetics Journals}

\IEEEtitleabstractindextext{%
\begin{abstract}

Magnetic skyrmions are emerging as potential candidates for next generation non-volatile memories. In this paper, we propose an in-memory binary neural network (BNN) accelerator based on the non-volatile skyrmionic memory, which we call as SIMBA. SIMBA consumes 26.7~mJ of energy and 2.7~ms of latency when running an inference on a VGG-like BNN. Furthermore, we demonstrate improvements in the performance of SIMBA by optimizing material parameters such as saturation magnetization, anisotropic energy and damping ratio. Finally, we show that the inference accuracy of BNNs is robust against the possible stochastic behavior of SIMBA (88.5\%~$\pm$1\%). 
\end{abstract}

}

\maketitle
\IEEEdisplaynontitleabstractindextext
\IEEEpeerreviewmaketitle

\input{sections/I/Sec_I}

\input{sections/II/Sec_II}

\input{sections/III/Sec_III}

\input{sections/IV/Sec_IV}

\input{sections/V/Sec_V}

\input{sections/VI/Sec_VI}

\section*{Acknowledgment}
This work was supported in part by Singapore Ministry of Education (MOE) Grant MOE 2017-T2-1-114, in part by MOE Academic Research Fund Tier 1, in part by the Agency for Science, Technology and Research (A*STAR) SpOT-LITE Program, in part by the NUS Start-up Grant, and in part by Competitive Research Programme (CRP) under Grant NRF-CRP12-2013-01.

\bibliographystyle{IEEEtran}
\bibliography{IEEEabrv,test.bib}

\vskip -2\baselineskip plus -1fil
\begin{IEEEbiography}[{\includegraphics[width=1in,height=1.25in,clip,keepaspectratio]{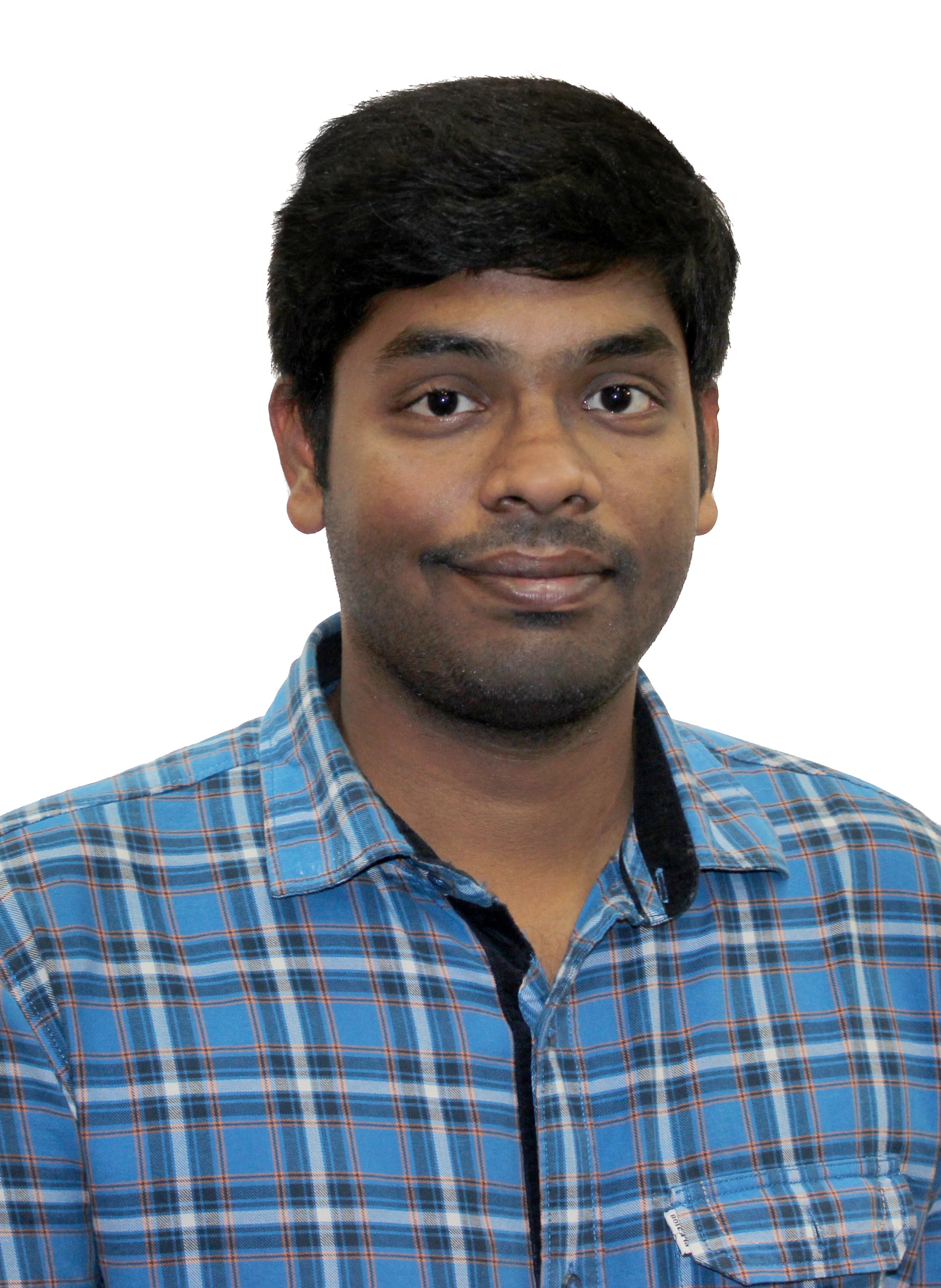}}]{Venkata Pavan Kumar Miriyala} received the B.Tech degree in electronics and communication engineering from SSN Engineering College, India in 2014. He previously held positions at IBM-Research Tokyo, Japan, and Indian Institute of Technology (IIT), Hyderabad. He is currently working towards a Ph.D. at the National University of Singapore. 
\end{IEEEbiography}
\vskip -2\baselineskip plus -1fil
\begin{IEEEbiography}[{\includegraphics[width=1in,height=1.25in,clip,keepaspectratio]{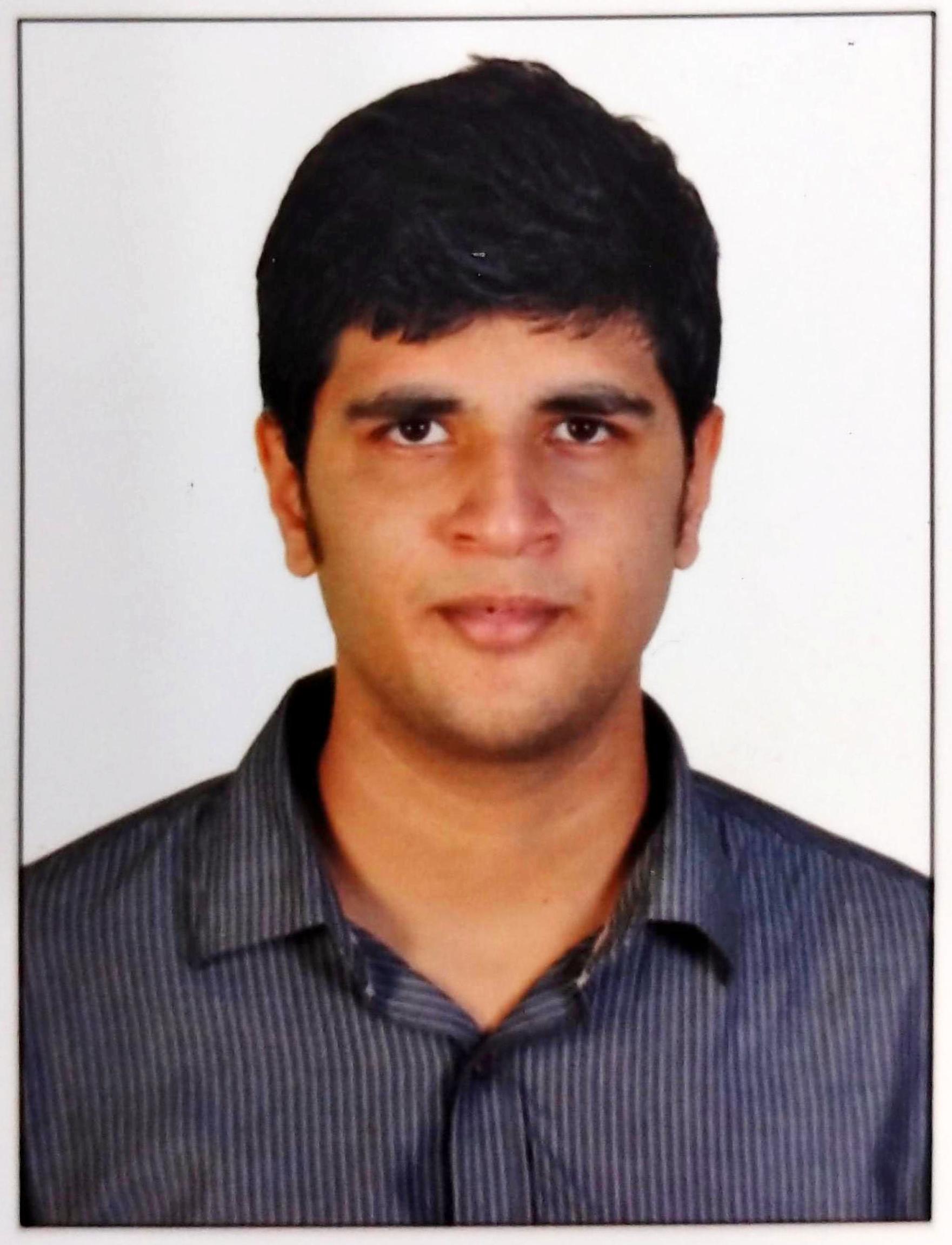}}]{Kale Rahul Vishwanath} received the M.Sc. Degree in Electrical Engineering from National University of Singapore and B.Tech. degree in Electronics Engineering from VJTI, Mumbai, India, in 2016 and 2014 respectively. He is currently working as a Research Engineer at the National University of Singapore.
\end{IEEEbiography}
\vskip -2\baselineskip plus -1fil
\begin{IEEEbiography}[{\includegraphics[width=1in,height=1.25in,clip,keepaspectratio]{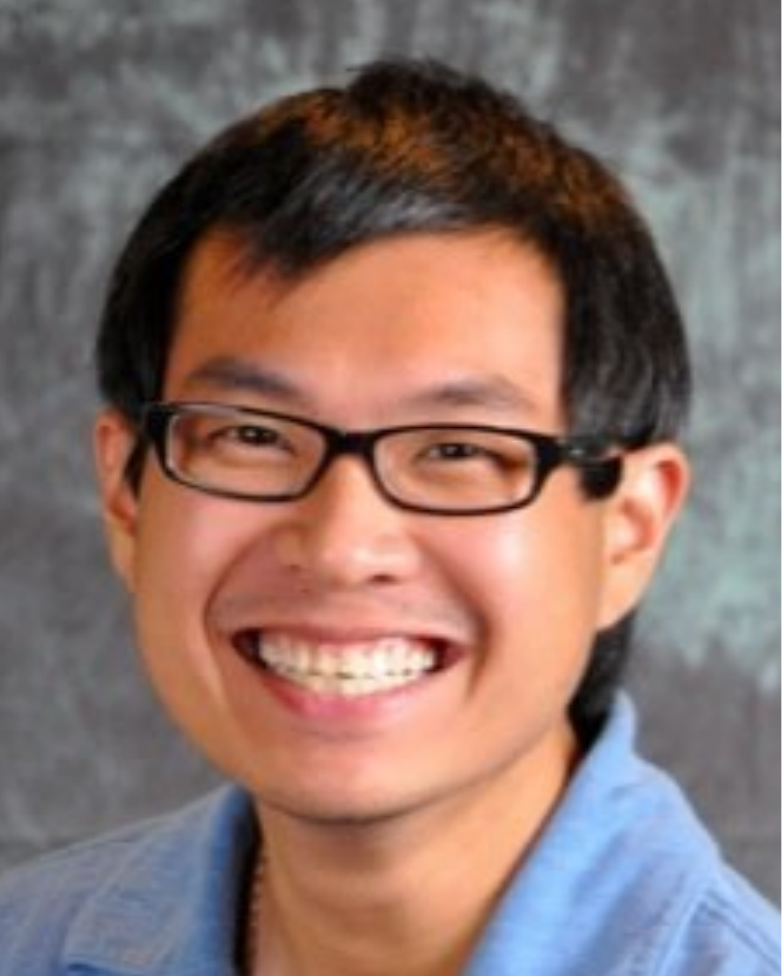}}]{Xuanyao Fong (SM$'$06, M$'$14)} received the Ph.D. and B.Sc. degrees in electrical engineering from Purdue University, West Lafayette, IN, in 2014 and 2006, respectively. He previously held positions at Advanced Micro Devices, Inc., Boston Design Center, Boxboro, MA, was a Research Assistant and Postdoctoral Research Assistant in Purdue University, and a Research Scientist in the Institute of Microelectronics, A*STAR, Singapore. Dr. Fong is currently an Assistant Professor in the Department of Electrical and Computer Engineering, National University of Singapore.
\end{IEEEbiography}
\end{document}

%% file: sections/I/Sec_I.tex
\section{Introduction}
\label{sec:intro}

Spintronic memories utilizing magnetic skyrmions have recently been the subject of great interest \cite{Fert2017, Chen2019, Roy2019, Lai2017, jiang2016, Tomasello2014}. This is due to scalability of skyrmions down to 2 nm \cite{Wang2018, Meyer2019, Romming2015}, the ease of electrical manipulation \cite{Romming2013, Fert2017} and their robustness and stability due to topological protection of skyrmions \cite{Sampaio2013, Nagaosa2013}. Existing skyrmionic memory proposals replace the magnetic domain wall devices in racetrack memories \cite{Parkin2008, Parkin2015} with their skyrmionic counterparts. However, micro-magnetic simulations recently show that when a spin-torque nano-oscillator (STNO) \cite{chen2016} is used to generate spin waves (SWs) in a ferromagnetic film, the presence of a skyrmion in the film can influence its steady state dynamics \cite{pavan2019}. A skyrmion is topologically protected and hence, does not allow the SWs to pass through it. Moreover, the skyrmion backscatters the SWs as if it is reradiating the SWs. The interference between SWs emitted by the STNO and SWs backscattered by the skyrmion, which are well-described by wave equations, can affect the steady state dynamics of the STNO---sufficiently strong constructive interference between SWs can lead to nucleation of a skyrmion in the STNO \cite{pavan2019}. 

In this paper, we propose a novel way of using the SW mediated interactions between magnetic skyrmions and STNOs to augment the skyrmionic memory with compute capabilities and implement an ultra-low power non-volatile in-memory computing engine (NVIMCE) using the presence and absence of a skyrmion as a state variable. To the best of our knowledge, this is the first skyrmionic in-memory computing proposal. Moreover, unlike other skyrmionic logic devices that have been proposed in the literature \cite{Meghna2019, Luo2018, Zhang2017}, the issues arising from the transverse motion of skyrmions \cite{Sampaio2013} are avoided in our proposed device because logic operations do not rely on skyrmion motion. 

On the other hand, the NVIMCE, which saves on energy associated with data storage and movement is a crucial design technique for accelerating neural network algorithms \cite{Chang2019, Fan2017, Chi2016}. The NVIMCEs can be powered down to achieve near-zero standby power consumption when the device is in sleep mode. When the device wakes up, data in the NVIMCE is updated and processed in-situ, which saves on latency and energy consumed to transfer data between memory and processing elements in the conventional von Neumann architecture. 

In this paper, we also propose a skyrmionic NVIMCE (abbreviated as SIMC) based hardware accelerator for binary neural networks (BNNs) \cite{Cour2016}, which we call as SIMBA. A hybrid simulation framework consisting of device, circuit and architecture-level simulation tools is developed to design, analyze and evaluate SIMBA. Our simulation results show that high energy efficiency and fast inference operations are achieved in SIMBA due to reduced data movement. Further, we have also demonstrated improvements in performance of SIMBA by optimizing the material parameters. 

The rest of this paper is structured as follows. The SW mediated interactions between skyrmions and STNOs, and the skyrmionic logic gates will be first introduced in Section~\ref{sec:Skylogic}. The hardware architectures of our proposed SIMC and SIMBA are then presented in Section~\ref{sec:SIMC} and Section~\ref{sec:SIMBA}, respectively. Results and discussion on proposed devices, circuits and architectures are then presented in Section~\ref{sec:Eval}. Finally, Section~\ref{sec:conclusion} concludes this paper.

\begin{figure*}[t]
\centering
\includegraphics[width=5in]{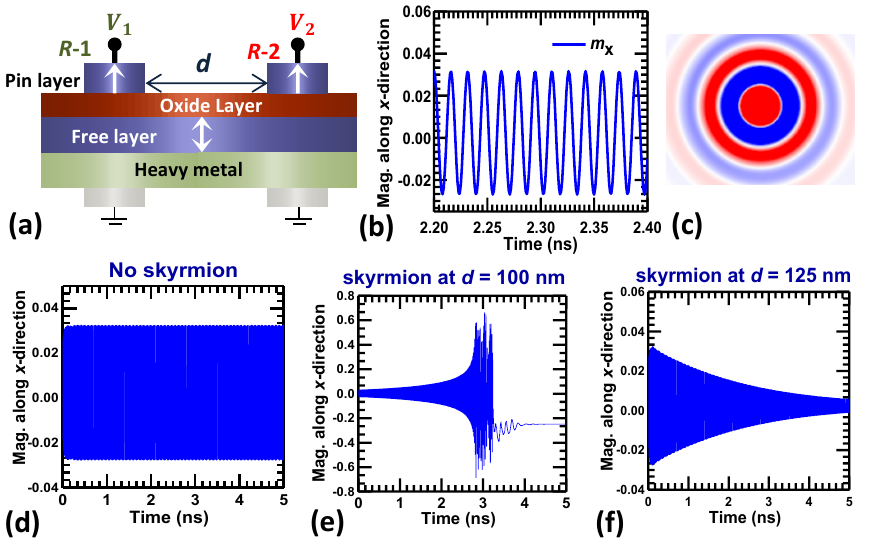}
\caption{The cross-sectional view of the device structure used to demonstrate the interactions between skyrmions and STNOs. $V_\text{1}$ generates the skyrmion in region $\it{R}$-1 and $V_\text{2}$ causes the local magnetization in region $\it{R}$-2 to precess continuously, shown in (b) and (c). The precession leads to the propagation of SWs outward from $\it{R}$-2 as shown in (c). (d) The local magnetization in $\it{R}$-2 when no skyrmion in present in $\it{R}$-1. (e) and (f) show the local magnetization in $\it{R}$-2 when a skyrmion is at $\textbf{\textit{d}}$ = 100~nm and $\textbf{\textit{d}}$ = 125~nm, respectively. The stabilized negative in-plane magnetization in (e) denotes the formation of new skyrmion in $\it{R}$-2.}
\label{fig:IntSS}
\addtocounter{figure}{0}
\end{figure*}

%% file: sections/II/Sec_II.tex
\section{Skyrmionic Logic Computation Using STNOs}
\label{sec:Skylogic}

For completeness sake, the underlying device physics of the non-volatile skyrmionic devices with which SIMBA is implemented is briefly presented in the following subsection. Further details of the device-level simulations are discussed in Section~\ref{sec:Eval}.

\begin{figure}[t]
\centering
\includegraphics[width=3.5 in]{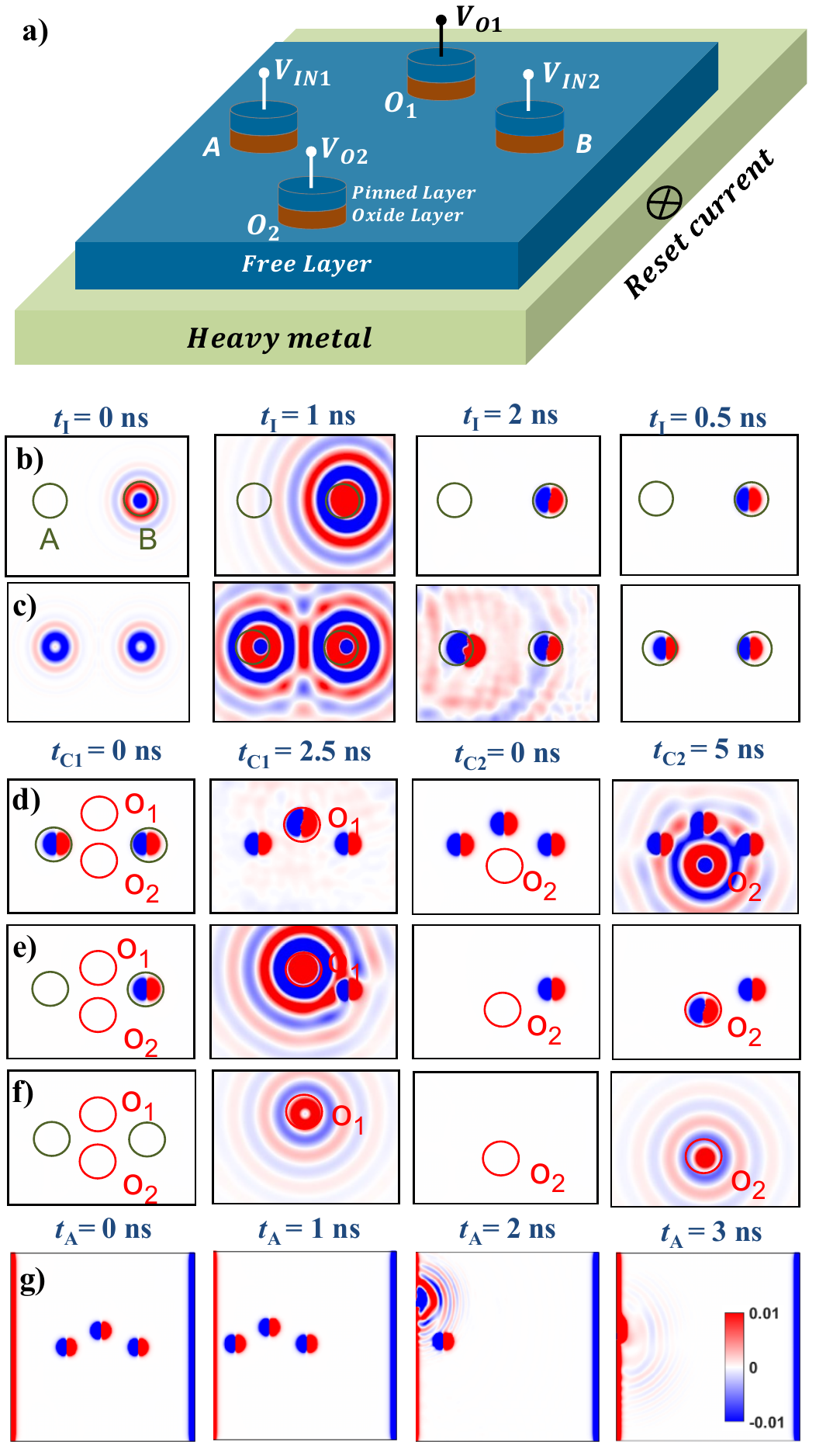}
\caption{(a) The proposed skyrmionic logic device. Regions, $\bm{A}$ and $\bm{B}$ are used for storing inputs whereas $\bm{{O}_\text{1}}$ and $\bm{{O}_\text{2}}$ are used for computing. Micromagnetic simulation of (b-c) input skyrmion nucleation, (d-f) logic computation and (g) skyrmion annihilation (reset) operations. The $\it{x}$-component of magnetization ($\bm{{m}_\text{x}}$) are shown in (b-g). The blue and red dumbbell-like structures in (d-g) are the skyrmions. (g) Skyrmions are annihilated by passing an in-plane reset current, illustrated in (a), through the HM, which pushes skyrmions to the boundary of the FL. The time delays during nucleation of inputs, computation during phase-1 and phase-2, and annihilation operations are denoted as ${t}_\text{I}$, ${t}_\text{C1}$, ${t}_\text{C2}$ and ${t}_\text{A}$, respectively. The inset in (g) shows the color map used for plotting $\bm{{m}_\text{x}}$ in (b-g).}
\label{fig:imcunit}
\addtocounter{figure}{0}
\end{figure}

\subsection{Interactions between Skyrmions and STNOs}
\label{subsec:swaminskystor}

Let us first consider the device structure shown in Fig.~\ref{fig:IntSS}~(a), which consists of an oxide/ferromagnetic free layer (FL)/heavy metal (HM) trilayer structure in which the magnetization of FL can be manipulated. Two magnetically pinned layers (PL) are placed on top of the oxide layer to form the magnetic tunnel junctions (MTJs) \cite{Ikeda2010, avan2019} labeled as $\it{R}$-1 and $\it{R}$-2.

Now, consider the injection of current into the FL through $\it{R}$-2 (by applying a voltage, ${V}_\text{2}$, across the corresponding MTJ) in the absence of any skyrmions in the FL layer. Results of the micromagnetic simulation in MuMax3 \cite{Arne2014}, shown in Fig.~\ref{fig:IntSS}~(b), demonstrate that current-induced spin-transfer torque (STT) \cite{Xiao2004} can cause the local magnetization in $\it{R}$-2 to oscillate continuously like an STNO. Due to the interaction between various magnetization energies (exchange, demagnetizing, anisotropies, etc.), the oscillations in $\it{R}$-2 radiate outward through the rest of FL in the form of SWs as Fig.~\ref{fig:IntSS}~(c) shows. Hence, applying a voltage across the MTJ in $\it{R}$-2 in this manner \emph{activates} the STNO at $\it{R}$-2.

\begin{table}[b!]
\centering
\caption{Simulation Parameters}
\includegraphics[width=3.5in, clip, trim=0.25in 0.17in 0.2in 0.17in, scale=1.0]{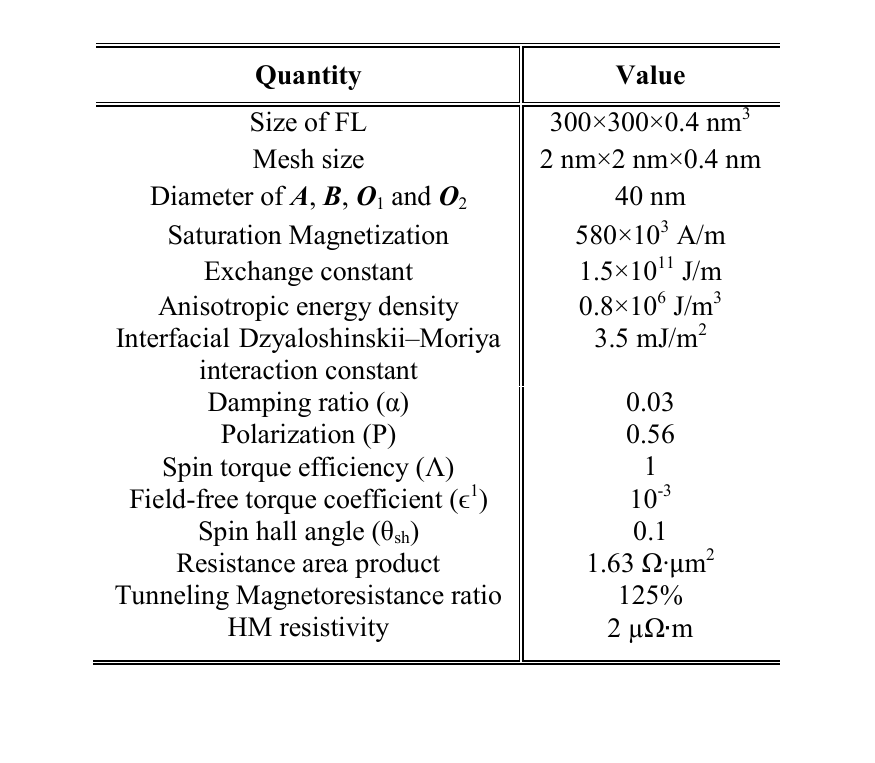}
\label{tab:sim_param}
\end{table}

However, if the current-induced STT on the FL is sufficiently strong, the local magnetization of the FL may switch to a skyrmionic magnetization state instead of continuously oscillating, as shown in Fig.~\ref{fig:IntSS}~(b). A skyrmion acts as a scatterer of SWs due to its topological protection. Hence, if a skyrmion is present at $\it{R}$-1 (which may be nucleated by applying a voltage, ${V}_\text{1}$, across the corresponding MTJ) before the STNO at $\it{R}$-2 is activated, the SWs emitted by the STNO at $\it{R}$-2 ($\it{SWs}_\text{emitted}$) will be backscattered by the skyrmion at $\it{R}$-1. The SWs backscattered by the skyrmion in $\it{R}$-1 ($\it{SWs}_\text{scattered}$) interfere with $\it{SWs}_\text{emitted}$ and can affect the dynamics of the STNO at $\it{R}$-2. Results of MuMax3 simulations show that STNO oscillations are amplified (Fig.~\ref{fig:IntSS}~(e)) and attenuated (Fig.~\ref{fig:IntSS}~(f)) when the interference is constructive and destructive, respectively. This phenomenon can be modeled using the following wave equations
\begin{eqnarray}
\label{eq:inter1}
\it{SWs}_\text{emitted} & \propto & A_\text{1} e^{\frac{k}{k_\text{f}}x-\omega t} cos(kx -\omega t)\\
\it{SWs}_\text{scattered} & \propto & B_\text{1} e^{\frac{k}{k_\text{f}}x-\omega t - \frac{\Phi}{2\Phi_\text{f}}} cos(kx -\omega t-\Phi)\\
\it{SWs}_\text{total} & = & \it{SWs}_\text{emitted} + \it{SWs}_\text{scattered}
\end{eqnarray}
where $A_\text{1}$ and $B_\text{1}$ are the wave amplitudes, $\it{k}$ = 2$\pi$/$\bm{\lambda}$ is the wave vector, $\bm{\lambda}$ is the wave length, $\omega$ is the angular frequency, $k_\text{f}$ and $\Phi_\text{f}$ are fitting constants, $\Phi$ = 2$\it{k}$($\textbf{\textit{d}}$-R) is the phase difference, and R is the radius of the skyrmion. When $\bm{\lambda}$ is constant, $\it{SWs}$$_\text{emitted}$ and $\it{SWs}$$_\text{scattered}$ interfere constructively for \textbf{\textit{d}} = $\bm{\lambda}$+ n$\bm{\lambda}$/4 $\forall$ n $\in$ {even positive integers}, whereas they interfere destructively for \textbf{\textit{d}} = $\bm{\lambda}$+ n$\bm{\lambda}$/4 $\forall$ n $\in$ {odd positive integers}. Interestingly, it was found that when the interfacial Dzyaloshinskii-Moriya interactions are sufficiently strong, the constructive interference leads to nucleation of a skyrmion in the STNO in $\it{R}$-2 \cite{pavan2019} (Fig.~\ref{fig:IntSS}~(e)). The stabilized negative $\it{x}$-component of magnetization in Fig.~\ref{fig:IntSS}~(e) denotes the nucleation of a skyrmion. We will next discuss the design of skyrmionic logic gates that utilize the effects discussed in this sub-section.

\begin{figure*}[t]
\centering
\includegraphics[width=7in]{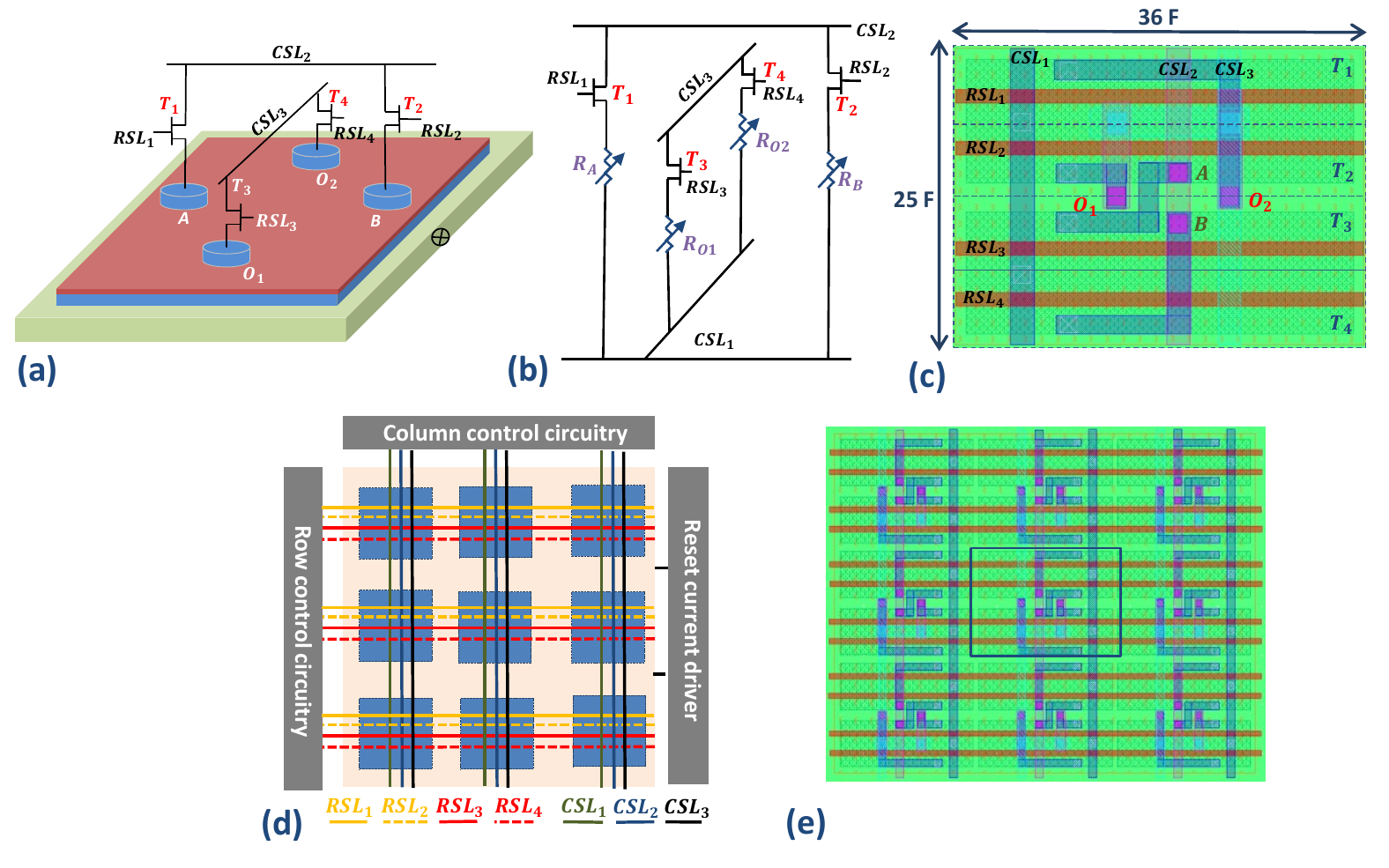}
\caption{(a) The device structure of a bit-cell in the proposed SIMC with four access transistors (${T}_\text{1}$, ${T}_\text{2}$, ${T}_\text{3}$ and ${T}_\text{4}$). Regions, $\bm{A}$, $\bm{B}$, $\bm{{O}_\text{1}}$ and $\bm{{O}_\text{2}}$ are connected to select line, ${CSL}_\text{1}$ at the bottom (not shown in (a) for better illustration). (b) Electrical model of the bit-cell. The regions $\bm{A}$, $\bm{B}$, $\bm{{O}_\text{1}}$ and $\bm{{O}_\text{2}}$ are assumed as simple MTJs with resistances, ${R}_\text{A}$, ${R}_\text{B}$, ${R}_\text{O1}$ and ${R}_\text{O2}$. Therefore, the proposed bit-cell can be seen as a simple circuit block with four 1-transistor 1-resistances. (c) Layout schematic of a bit-cell obtained using the NCSU FreePDK45 \cite{NCSU2011}. (d) Array-level design of SIMC with 3$\times$3 bit-cells. Row select lines, ${RSL}_\text{1}$, ${RSL}_\text{2}$ are represented as yellow-colored line and dotted-lines, respectively. Similarly, ${RSL}_\text{3}$, ${RSL}_\text{4}$ are represented as red-colored line and dotted-lines, respectively. The column select lines, ${CSL}_\text{1}$, ${CSL}_\text{2}$ and ${CSL}_\text{3}$ are represented as green, blue and black-colored lines, respectively. (e) An example layout of this 3$\times$3 bit-cell array. The rectangular box in (e) marks a single bit-cell.}
\label{fig:layout}
\addtocounter{figure}{0}
\end{figure*}

\subsection{Spin-wave-based Skyrmionic Logic}
\label{subsec:Skylogic}

The structure of our proposed skyrmionic logic gate, shown in Fig.~\ref{fig:imcunit}~(a), contains four point-contact regions ($\bm{A}$, $\bm{B}$, $\bm{{O}_\text{1}}$ and $\bm{{O}_\text{2}}$). The regions $\bm{A}$ and $\bm{B}$ are used for storing inputs, whereas regions $\bm{{O}_\text{1}}$ and $\bm{{O}_\text{2}}$ are used during computation. The presence (absence) of the skyrmions in these regions denote the bit `1' (`0') and can be electrically sensed as the resistance of corresponding MTJs \cite{Ikeda2010}. Fig.~\ref{fig:imcunit}~(b) shows the time evolution of FL magnetization when $\bm{A}$ and $\bm{B}$ are programmed to be `0' and `1', respectively, whereas Fig.~\ref{fig:imcunit}~(c) shows the time evolution of FL magnetization when $\bm{A}$ and $\bm{B}$ are both programmed to be `1'. 

Computation on inputs stored at $\bm{A}$ and $\bm{B}$ is performed by using $\bm{{O}_\text{1}}$ and $\bm{{O}_\text{2}}$ as STNOs. $\bm{{O}_\text{1}}$ and $\bm{{O}_\text{2}}$ are placed at distance, \textbf{\textit{d}}~=~125~nm apart from each other and at a distance, \textbf{\textit{d}}~=~150~nm from both the input bits, $\bm{A}$ and $\bm{B}$. The material parameters assumed in this work lead to $\bm{\lambda}~\approx~100$~nm. In this geometry, if the STNO at $\bm{{O}_\text{1}}$ is activated, the presence of a skyrmion at either $\bm{A}$ or $\bm{B}$ leads to constructive interference between $\it{SWs}_\text{emitted}$ and $\it{SWs}_\text{scattered}$ in $\bm{{O}_\text{1}}$ and amplify the oscillations in the STNO, resulting in the formation of a skyrmion at $\bm{{O}_\text{1}}$. Furthermore, if skyrmions are present in both $\bm{A}$ and $\bm{B}$, the amplification in STNO oscillations would be large and a skyrmion forms at $\bm{{O}_\text{1}}$ within relatively short time delays. Due to symmetry, the same effect is observed when the STNO in $\bm{{O}_\text{2}}$ is activated. However, in the presence of a skyrmion at $\bm{{O}_\text{1}}$, $\it{SWs}_\text{emitted}$ by $\bm{{O}_\text{2}}$ and $\it{SWs}_\text{scattered}$ by the skyrmion $\bm{{O}_\text{1}}$ interfere destructively and oscillations at $\bm{{O}_\text{2}}$ become attenuated.

Consider only the STNO at $\bm{{O}_\text{1}}$ with no skyrmions in $\bm{{O}_\text{2}}$. If $\bm{{O}_\text{1}}$ is activated for 2.5~ns, a new skyrmion will only form in $\bm{{O}_\text{1}}$ if skyrmions are present in both $\bm{A}$ and $\bm{B}$ (see Fig.~\ref{fig:imcunit}~(d) from ${t}_\text{C1}$ = 0~ns to 2.5~ns). Otherwise, no skyrmion will form in $\bm{{O}_\text{1}}$ (see Fig.~\ref{fig:imcunit}~(e)-(f) from ${t}_\text{C1}$ = 0~ns to 2.5~ns). Hence,  the output stored at $\bm{{O}_\text{1}}$ may be expressed using the Boolean expression $\bm{{O}_\text{1}}$~=~$\bm{A}\cdot\bm{B}$ (\emph{i.e.}, $\bm{A}$~$\it{AND}$~$\bm{B}$). If only one of $\bm{A}$ or $\bm{B}$ has a skyrmion instead, the pulse width of ${V}_\text{O1}$ needs to be 5~ns in order for a skyrmion to form in $\bm{{O}_\text{1}}$. Due to the symmetry between $\bm{{O}_\text{1}}$ and $\bm{{O}_\text{2}}$, same effect can be observed at $\bm{{O}_\text{2}}$ as well. We demonstrate the latter case by activating $\bm{{O}_\text{2}}$ for 5~ns (see Fig.~\ref{fig:imcunit}~(e)-(f) from ${t}_\text{C2}$ = 0~ns to 5~ns). The output stored at $\bm{{O}_\text{2}}$ may be expressed using the Boolean expression $\bm{{O}_\text{2}}$~=~$\bm{A}+\bm{B}$ (\emph{i.e.}, $\bm{A}$~$\it{OR}$~$\bm{B}$). Hence, $\bm{A}\cdot\bm{B}$ and $\bm{A}+\bm{B}$ operations can be performed by ensuring that $\bm{{O}_\text{2}}$ and $\bm{{O}_\text{1}}$ do not have any skyrmion before activating the STNO with the required pulse width at the desired location.

The $\it{XOR}$ operation between $\bm{A}$ and $\bm{B}$ ($\bm{A}$~$\oplus$~$\bm{B}$) needs to be performed in two phases. First, we note that the Boolean expression for the $\it{XOR}$ operation may be written as
\begin{equation}
\label{eq:boolean}
\bm{A}~\oplus~\bm{B} = (\bm{A}~\cdot~\overline{\bm{B}}~+~\overline{\bm{A}}~\cdot~\bm{B}) = (\bm{A}~+~\bm{B})~\cdot~(\overline{\bm{A}~\cdot~\bm{B}})
\end{equation}
Hence, in the first phase, $\bm{A}$~$\cdot$~$\bm{B}$ operation is performed and stored at $\bm{{O}_\text{1}}$. In the second phase, the voltage pulse width is given to the STNO in $\bm{{O}_\text{2}}$ as if $\bm{A}+\bm{B}$ operation is being performed. However, the result of the output of $\bm{{O}_\text{2}}$ depends on the output stored in $\bm{{O}_\text{1}}$, as discussed earlier. Due to destructive interference between $\it{SWs}_\text{emitted}$ by $\bm{{O}_\text{2}}$ and $\it{SWs}_\text{scattered}$ by the skyrmion $\bm{{O}_\text{1}}$, the result at $\bm{{O}_\text{2}}$ at the end of the two-phase operation is $(\bm{A}+\bm{B})\cdot(\overline{\bm{A}\cdot\bm{B}})~=~\bm{A}~\oplus~\bm{B}$. For instance, if no skyrmions are present in both $\bm{A}$ and $\bm{B}$ (\emph{i.e.}, $\bm{A}$ = $\bm{B}$ = `0'), magnetization at $\bm{{O}_\text{1}}$ during first phase and magnetization at $\bm{{O}_\text{2}}$ during second phase would only oscillate continuously and no skyrmion will form at $\bm{{O}_\text{2}}$, which is in accordance with the Eq.~\ref{eq:boolean}. If a skyrmion is present at either $\bm{A}$ or $\bm{B}$ (\emph{i.e.}, $\bm{A}~\oplus~\bm{B}$ = `1'), the $\it{SWs}_\text{emitted}$ by $\bm{{O}_\text{2}}$ and $\it{SWs}_\text{scattered}$ by the skyrmion interfere constructively at $\bm{{O}_\text{2}}$. Consequently, a skyrmion will form at $\bm{{O}_\text{2}}$ in agreement with the Eq.~\ref{eq:boolean}. Finally, if skyrmions are present at both $\bm{A}$ and $\bm{B}$ (\emph{i.e.}, $\bm{A}$~=~$\bm{B}$ = `1'), they would also cause the SWs to interfere constructively at $\bm{{O}_\text{2}}$. However, the interference between SWs at $\bm{{O}_\text{2}}$ is not as strong in the presence of SWs backscattered by the skyrmion that is nucleated at $\bm{{O}_\text{1}}$ during the first phase. Consequently, a skyrmion will \emph{not} form in $\bm{{O}_\text{2}}$ when skyrmions are present in \emph{both} $\bm{A}$ and $\bm{B}$, which satisfies the Eq.~\ref{eq:boolean}. The $\it{XOR}$ operation between $\bm{A}$ and $\bm{B}$ operation is validated in micromagnetic simulations shown in Fig.~\ref{fig:imcunit}~(d)-(f). As shown in Fig.~\ref{fig:imcunit}~(d), when $\bm{A}$ and $\bm{B}$ are `1', $\bm{{O}_\text{2}}$ is `0'. If only one of $\bm{A}$ or $\bm{B}$ = `1', $\bm{{O}_\text{2}}$ = `1' (Fig.~\ref{fig:imcunit}~(e)). Finally, when both $\bm{A}$ and $\bm{B}$ are `0', $\bm{{O}_\text{2}}$ is `0' (Fig.~\ref{fig:imcunit}~(f)).   

Note that if any skyrmion in the FL is to be annihilated in any operation, then all skyrmions in the FL would have to be annihilated. Skyrmions are then nucleated at the desired locations after the annihilation (reset) process. The reset process is performed by passing current in the plane of the HM layer. A spin-orbit torque (SOT) induced by the current flow is exerted on the FL magnetization \cite{Liu2012}. If this torque is sufficiently strong, skyrmions get pushed off the boundaries of the FL. Simulation results in Fig.~\ref{fig:imcunit}~(g) show the scenario where skyrmions are getting pushed towards the left boundary of the FL and finally getting annihilated.

%% file: sections/III/Sec_III.tex
\section{Skyrmionic In-memory Computing Engine (SIMC)}
\label{sec:SIMC}

The device structure presented in Section~\ref{subsec:Skylogic} is used to implement a multi-level bit-cell (see Fig.~\ref{fig:layout}) in our proposed SIMC. Since our proposed device is non-volatile, it may also be used as a memory device that stores data at $\bm{A}$ and $\bm{B}$. As discussed in Section~\ref{subsec:Skylogic}, $\bm{{O}_\text{1}}$ and $\bm{{O}_\text{2}}$ may be used for performing logic operations on the stored data.

\subsection{Multi-level bit cell}
\label{subsec:bitcell}

Each of the four regions of our proposed device is connected to an access transistor, as Fig.~\ref{fig:layout}~(a) shows. Regions $\bm{A}$ and $\bm{B}$ are connected to column select line, ${CSL}_\text{1}$ at the bottom, and to column select line, ${CSL}_\text{2}$ through transistors, ${T}_\text{1}$ and ${T}_\text{2}$ (which are controlled by row select lines, ${RSL}_\text{1}$ and ${RSL}_\text{2}$, respectively). Similarly, regions $\bm{{O}_\text{1}}$ and $\bm{{O}_\text{2}}$ are connected to column select line, ${CSL}_\text{1}$ at the bottom, and to column select line, ${CSL}_\text{3}$ through transistors, ${T}_\text{3}$ and ${T}_\text{4}$ (which are controlled by row select lines, ${RSL}_\text{3}$ and ${RSL}_\text{4}$, respectively). For better illustration, connections to ${CSL}_\text{1}$ at the bottom are omitted from Fig.~\ref{fig:layout}~(a). However, all the connections are shown in Fig.~\ref{fig:layout}~(b), which represents the circuit model of our proposed bit-cell.

The write operation to the bit-cell occurs as follows. When bit `1' is to be written into $\bm{A}$ (\emph{i.e.}, nucleate a skyrmion in $\bm{A}$), ${T}_\text{1}$ is turned ON to pass current from ${CSL}_\text{2}$ to ${CSL}_\text{1}$ through $\bm{A}$. To write bit `0' into $\bm{A}$ (\emph{i.e.}, do not nucleate a skyrmion), ${T}_\text{1}$ is turned OFF and no current passes through $\bm{A}$. Similarly, to write bit `1' into $\bm{B}$, ${T}_\text{2}$ is turned ON and the current is passed from ${CSL}_\text{2}$ to ${CSL}_\text{1}$ through $\bm{B}$. To write bit `0' into $\bm{B}$, ${T}_\text{2}$ is turned OFF and no current is passed through $\bm{B}$. 

For performing the logic computations on data stored in $\bm{A}$ and $\bm{B}$, $\bm{{O}_\text{1}}$ and $\bm{{O}_\text{2}}$ are utilized as STNOs. To activate the STNO in $\bm{{O}_\text{1}}$, ${T}_\text{3}$ is turned ON and the current is passed from ${CSL}_\text{3}$ to ${CSL}_\text{1}$ through ${T}_\text{3}$. Otherwise, ${T}_\text{3}$ is turned OFF and no current is passed through $\bm{{O}_\text{1}}$. Similarly, to activate the STNO in $\bm{{O}_\text{2}}$, ${T}_\text{4}$ is turned ON and the current is passed from ${CSL}_\text{3}$ to ${CSL}_\text{1}$ through ${T}_\text{4}$. Otherwise, ${T}_\text{4}$ is turned OFF and no current is passed through $\bm{{O}_\text{2}}$.

Note that, as mentioned earlier in Section~\ref{subsec:Skylogic}, if any input bit needs to switch from `1' to `0' or if the logic computation needs to be re-evaluated, the reset current must first be passed in-plane through the HM to annihilate the skyrmions, followed by nucleation of skyrmions at the desired locations. This process may require the read operation of the data stored in $\bm{A}$ and $\bm{B}$. The data stored in $\bm{A}$ and $\bm{B}$, and the output of computation stored in $\bm{{O}_\text{1}}$ and $\bm{{O}_\text{2}}$, may be sensed as the resistance of the MTJ at the corresponding regions. A current-sensing scheme with conventional sense amplifiers may be used to perform the read operation \cite{Dong2012, Kim2015, Kim2014, Trinh2018}. Note that the voltages across the MTJs during read operations need to be optimized to avoid the read-failure or read-disturb errors.

\subsection{Hardware architecture of SIMC}
\label{subsec:HASIMC}

The proposed multi-level bit-cells are arranged into regular arrays to construct the SIMC. The layout for an individual bit-cell is depicted in Fig.~\ref{fig:layout}~(c). The ${CSL}_\text{1}$ and ${CSL}_\text{2}$ are routed on metal-2 whereas ${CSL}_\text{3}$ is routed on metal-3. The bit-cells are arranged into a large array and a 3$\times$3 area of the bit-cell array is shown in Fig.~\ref{fig:layout}~(d). The light orange-colored region in Fig.~\ref{fig:layout}~(d) depicts the HM layer that is shared among the bit-cells whereas the blue-colored square regions depict the magnetic device structures. The colored-lines and dotted-lines represent the various control lines to the bit-cells. Note that separate row and column control circuitry drive the control lines as shown in Fig.~\ref{fig:layout}~(d). Furthermore, a constant current driver supplies the reset current to the HM during reset operations. The layout of a 3$\times$3 area of the bit cell array is shown in Fig.~\ref{fig:layout}~(e).

\begin{figure*}[t]
\centering
\includegraphics[width=6in]{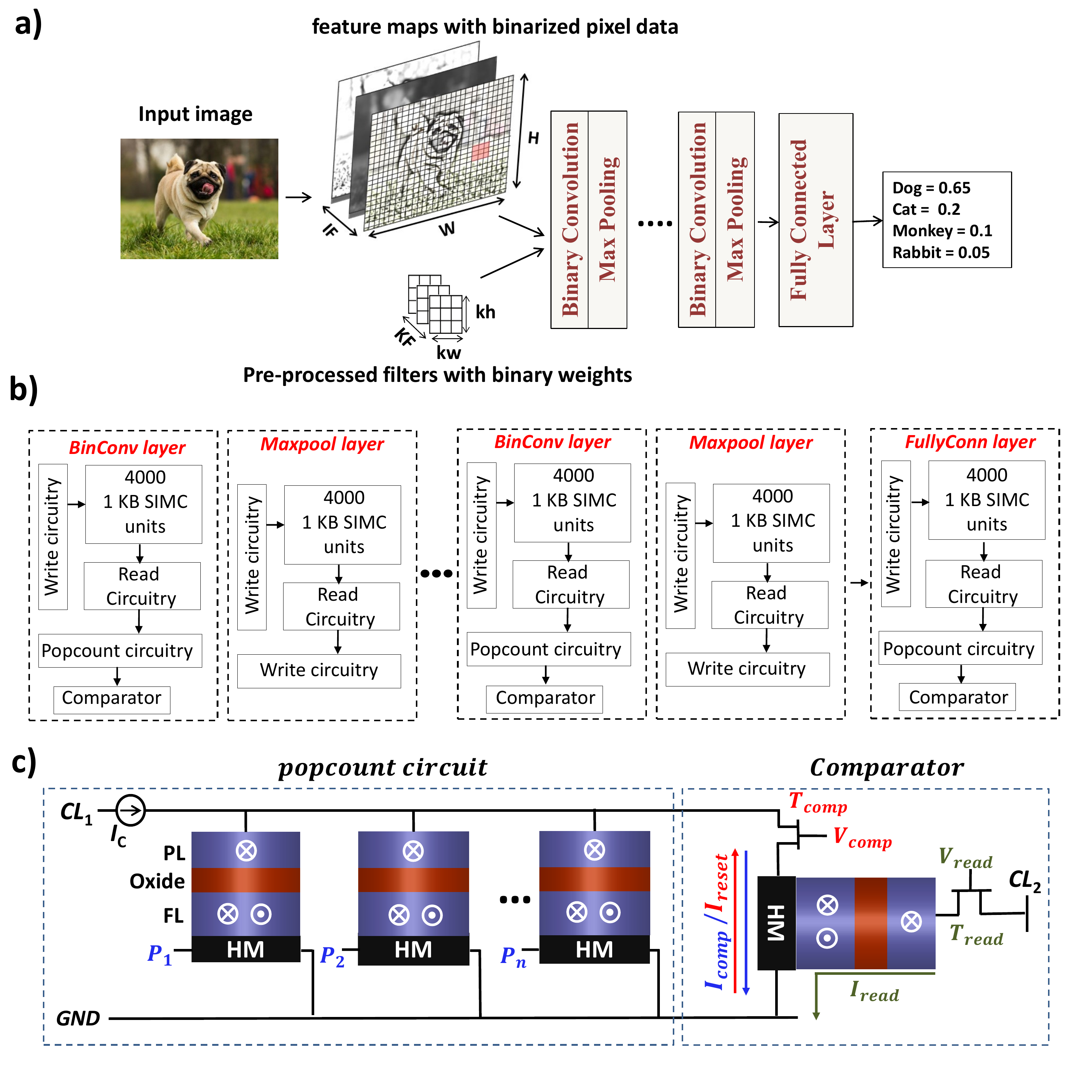}
\caption{(a) The data flow of BNNs computation when classifying an input image into different categories. A typical BNN consists of different binary convolution (BinConv), max pooling (Maxpool) and fully connected (FullyConn) layers. (b) SIMBA, our proposed hardware architecture for accelerating the BNN algorithm. 1~KB SIMC units are used to perform the 2-bit $\it{XOR}$/$\it{OR}$/$\it{AND}$ operations. (c) The proposed $\it{popcount}$ and $\it{comparator}$ circuits utilizing the spin-orbit torque MTJs.}
\label{fig:SIMBA}
\addtocounter{figure}{0}
\end{figure*}

%% file: sections/IV/Sec_IV.tex
\section{Hardware Architecture of SIMBA}
\label{sec:SIMBA}

The SIMC proposed in Section~\ref{sec:SIMC} can be used to implement the hardware accelerator for binary neural network (BNN) algorithms. We first introduce BNNs and show the operations that are needed to be performed. The mapping of these operations to dedicated hardware are then discussed before showing how they are put together to give the hardware implementation of SIMBA.

\subsection{Binary Neural Networks}
\label{sec:BNNs}

BNNs are a class of deep learning algorithms that perform inference operations using binarized inputs and weights \cite{Cour2016}. The binarized pixel data of images are first fed to the binary convolution (BinConv) layer in the form of input feature maps ($\it{IFs}$) with size $\it{(W, H)}$ (see Fig.~\ref{fig:SIMBA}~(a)). In BinConv layer, the data in $\it{IFs}$ is convolved with binary weights stored in filters ($\it{KFs}$) of size $\it{(kw, kh)}$. Mathematically,
\begin{eqnarray}
\begin{aligned}
\label{eq:binconv1}
\it{f(i,j)} & = \it{popcount}(IF(W,H)~~\it{XNOR}~~KF(kw,kh)) \\
\it{OF(i, j)} & = 1; \text{if}~\it{f(i,j)} \geq \it{threshold(f)} \\
& = 0; \text{if}~\it{f(i,j)} < \it{threshold(f)}
\end{aligned}
\end{eqnarray}
where ($\it{i,j}$) are the array indices of output feature map ($\it{OF}$). When given a string of bits, the $\it{popcount}$ operation counts the number of `1's in the string. As expressed in Eq.~\ref{eq:binconv1}, convolution on binary data is achieved by performing a sequence of bit-wise $\it{XNOR}$, $\it{popcount}$ and integer $\it{comparator}$ operations.

The dimensions of resultant $\it{OFs}$ from the BinConv layer are compressed in the maxpooling (Maxpool) layer. In the Maxpool layer, the maximum pixel value in a selected group of pixel values is estimated and forwarded to the subsequent layers. Due to the binarization of the data, the results of the Maxpool layer can be obtained using bit-wise $\it{OR}$ operations. As shown in Fig.~\ref{fig:SIMBA} (a), the compressed $\it{OFs}$ are propagated through several BinConv and Maxpool layers. The final $\it{OFs}$ are then given as inputs to the fully-connected (FullyConn) layer to obtain the inference/classification results. FullyConn layer is a feed-forward multi-perceptron network, where the weighted sum of inputs is compared with a threshold in each perceptron. Due to the binarization of the network, computations in this layer can also be executed using bit-wise $\it{AND}$ operations followed by $\it{popcount}$ and $\it{comparator}$.

\subsection{Hardware Implementation of SIMBA}
\label{sec:IMSIMBA}

The proposed hardware architecture of SIMBA is shown in Fig.~\ref{fig:SIMBA}~(b). It is assumed that training of the BNN is performed offline and the configuration of the trained BNN then is loaded into the SIMBA hardware, which performs the test/inference operations. The hardware architecture of SIMBA consists of hardware dedicated to the execution of operations for the BinConv, Maxpool, and FullyConn layers. Note from Eq.~\ref{eq:binconv1} that the BNN algorithm requires the computation of bit-wise $\it{XNOR}$ followed by $\it{popcount}$, which counts the number of bits that \emph{match} each other. The same result may be obtained by counting the number of bit \emph{mismatches} and subtracting the result from the total number of bit comparisons, which is known \emph{a priori}. Hence, the result of $f(i,j)$ in Eq.~\ref{eq:binconv1} can also be computed using $\it{XOR}$ operations (which returns whether the input bits \emph{mismatch}) followed by a suitably implemented $\it{popcount}$ circuit. For executing BinConv layers in SIMBA, SIMC units capable of 1~KB data storage perform the $\it{XOR}$ computations, which are read out and passed to the $\it{popcount}$ circuitry.

Since our SIMC units perform the $\it{XOR}$ operation, $OF(i,j)$ may be calculated using one of two methods. In the first method, the conventional $\it{popcount}$ is used to count the number of `1's returned by the $\it{XOR}$ operations. $OF(i,j)$ is set to `0' if the result exceeds the threshold and `1' otherwise. In the second method, a $\it{popcount}$ circuit that counts the number of `0's instead of `1' is given the result of the $\it{XOR}$ operations. $OF(i,j)$ is set to `1' if the result of exceeds the threshold and `0' otherwise. In this paper, we implemented the former method to compute $OF(i,j)$. Fig.~\ref{fig:SIMBA}~(c) shows our proposed $\it{popcount}$ circuit, which consists of `$\it{n}$' spin-orbit torque MTJs (SOT-MTJs), each of which receives as input the output of each $\it{XOR}$ operation. All the $\it{popcount}$ SOT-MTJs are initially set to low resistance state (\emph{i.e.}, state `0') by providing the negative reset voltage ($-V_\text{P}~<~0~\text{V}$) to the HMs of SOT-MTJs. The $\it{comparator}$ SOT-MTJ is initialized to the high resistance state (`1') by passing current through the HM of $\it{comparator}$ SOT-MTJ. When the output of $i$-th $\it{XOR}$ operation is `0' (`1'), $P_\text{i}$ is set as 0~V ($+V_\text{P}~>~0~\text{V}$). Positive voltage on $P_\text{i}$ causes current to flow through the corresponding HM, which switches the resistance state of the corresponding SOT-MTJ from low (\emph{i.e.}, state `0') to high (\emph{i.e.}, state `1'). After all $\it{XOR}$ results have been processed, the transistor, $T_\text{comp}$, is turned ON to pass positive current from $CL_\text{1}$ to the ground. The resistance states of $\it{popcount}$ MTJs determine the magnitude of $\it{I}_\text{comp}$ passing through the $\it{comparator}$ SOT-MTJ (see Fig.~\ref{fig:SIMBA}~(c)). If the magnitude of $\it{I}_\text{comp}$ exceeds the threshold switching current, the resistance state of the $\it{comparator}$ SOT-MTJ changes from high to low (`1' to `0'). The output result can read by turning ON the transistor, $T_\text{read}$ (by setting the read voltage $V_\text{read}$) and sensing the current flowing into the $\it{comparator}$ SOT-MTJ through $CL_\text{2}$.

For executing the Maxpool layer, as depicted in Fig.~\ref{fig:SIMBA}~(b), the same SIMC unit hardware design may be used. However, the SIMC units for Maxpool layers are configured to perform only $\it{OR}$ operations. Furthermore, $\it{popcount}$ and $\it{comparator}$ circuitry are not needed. Also shown in Fig.~\ref{fig:SIMBA}~(b), the hardware used for executing the computations in the FullyConn layer is similar to the hardware for executing the BinConv layer. However, the SIMC units for the FullyConn layer are configured to perform $\it{AND}$ operations only.

%% file: sections/V/Sec_V.tex
\section{Results and Discussion}
\label{sec:Eval}

We will now focus on the design and evaluation of the devices, circuits and hardware architecture that constitute SIMBA. In order to study the interactions between the different levels of design abstraction, we developed a hybrid mixed-mode devices-to-hardware architecture simulation framework. We will first present the simulation framework used for this work before discussing the simulation results for the devices, circuits and hardware architecture of SIMBA.

\begin{figure}[t]
\centering
\includegraphics[width=3.5in]{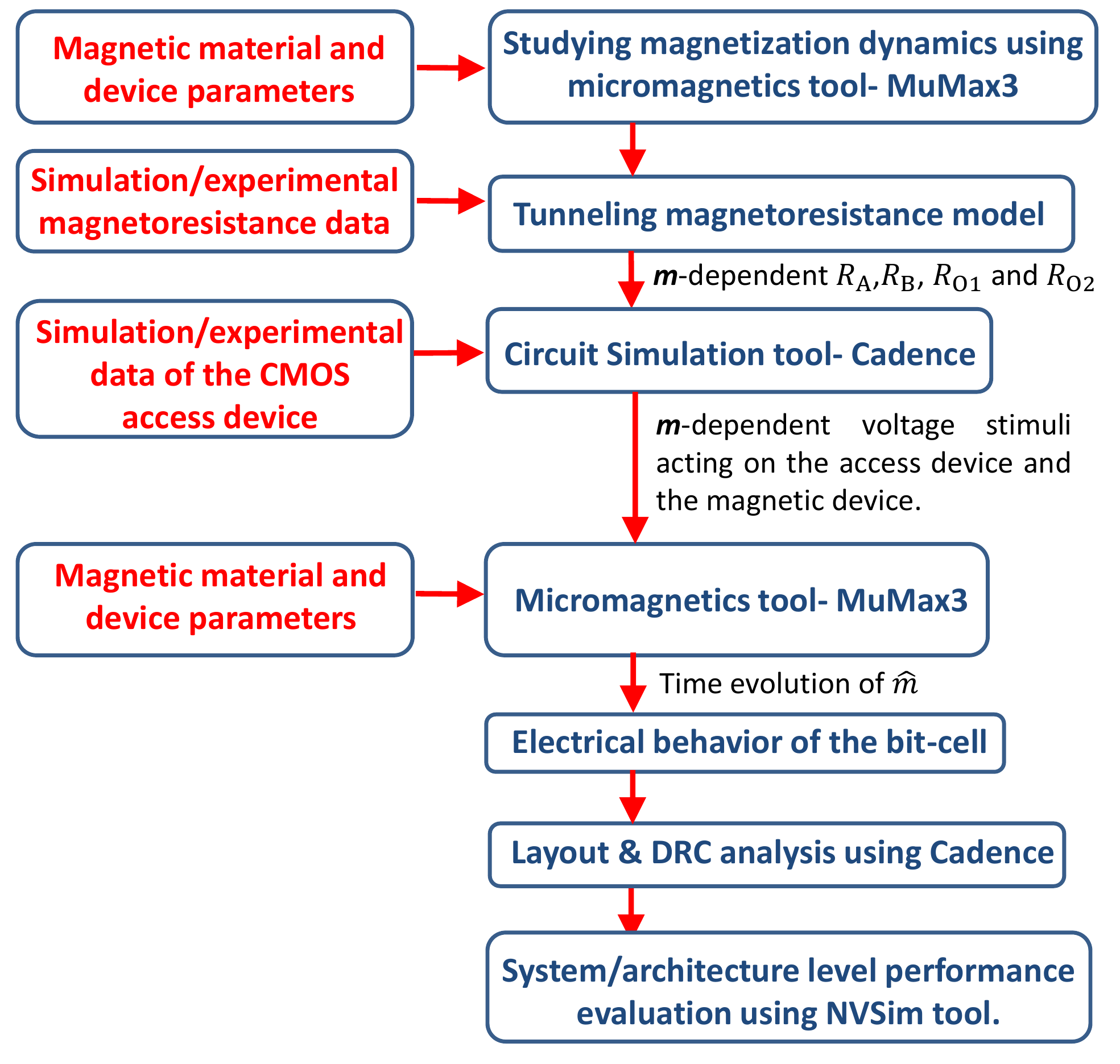}
\caption{Flow chart of the proposed simulation framework.}
\label{fig:flowchart}
\addtocounter{figure}{0}
\end{figure}

\begin{table*}[t]
\centering
\caption{Electrical inputs for the bit-cell during its different operations}
\includegraphics[width=5in, clip, scale=1.0]{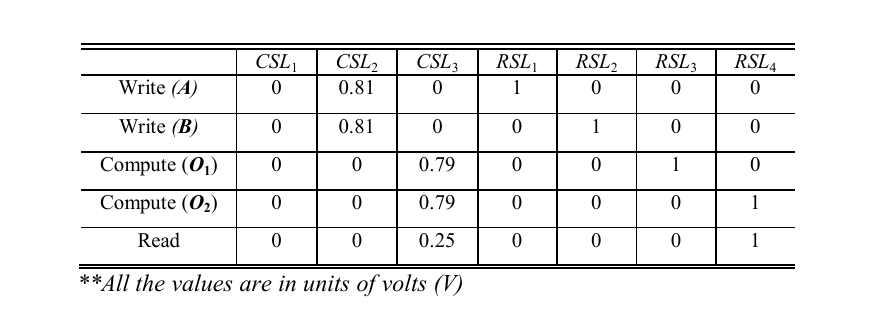}
\label{tab:inputs}
\end{table*}
\subsection{Modeling and Simulation Framework}
\label{subsec:SF}

Fig.~\ref{fig:flowchart} shows the flow-chart of our simulation framework. At the device-level, we used the MuMax3 micromagnetic simulator \cite{Arne2014} to study the magnetization dynamics and the interactions between skyrmions and STNOs. The magnetic device and material parameters we assumed in our simulations are extracted from experimental results \cite{Sampaio2013} and listed in Table~\ref{tab:sim_param}. We have modified the MuMax3 simulator \cite{Fong} to model the current flowing through an MTJ (in Fig.~\ref{fig:IntSS} and Fig.~\ref{fig:imcunit}) and the current-induced spin-transfer torque when a voltage is applied across it. 

The structure and electrical model of our proposed multi-level bit-cell are shown in Fig.~\ref{fig:layout}~(a)-(b). The resistances, ${R}_\text{A}$, ${R}_\text{B}$, ${R}_\text{O1}$ and ${R}_\text{O2}$ vary with respect to the magnetization state of the regions $\bm{A}$, $\bm{B}$, $\bm{O}_\text{1}$ and $\bm{O}_\text{2}$, respectively and can be written as 
\begin{eqnarray}
R_\text{X} & = & R_\text{P} + ({R_\text{AP}-R_\text{P}})\left(\frac{1-\bm{m}\cdot\bm{m_\text{p}}}{2}\right)
\label{eq:resistance}
\end{eqnarray}
where, $R_\text{X}$ $\in$ \{${R}_\text{A}$, ${R}_\text{B}$, ${R}_\text{O1}$ and ${R}_\text{O2}$\}, $\bm{m}$ and $\bm{m_\text{p}}$ are unit vectors describing the magnetization direction of FL and PL, respectively, in the corresponding regions. $R_\text{P}$ ($R_\text{AP}$) is the resistance of the region when $\bm{m}$ and $\bm{m_\text{p}}$ are parallel (anti-parallel) to each other. 

In order to study the magnetization dynamics at the bit-cell level, $\bm{m}$-dependent values of ${R}_\text{A}$, ${R}_\text{B}$, ${R}_\text{O1}$ and ${R}_\text{O2}$ are first calculated using Eq.~\ref{eq:resistance}. The electrical model of our proposed bit-cell is simulated using SPICE-like \cite{Spice2019, Cadence2019} circuit simulation tools to extract the electrical stimuli on the individual resistors and access transistors (Fig.~\ref{fig:layout}~(b)). In this work, we have modeled the access transistors using the NCSU FreePDK45 \cite{NCSU2011}. The circuit simulation results are used in the MuMax3 micromagnetic simulation to determine the magnetization dynamics. At this point, the characterization data about the individual bit-cell is available and fed to the array/architecture-level simulator. We have used the NVSim simulation tool \cite{Dong2012} to estimate the array/architecture-level energy consumption and latency of the proposed SIMC.

\begin{figure}[b]
\centering
\includegraphics[width=3.5in]{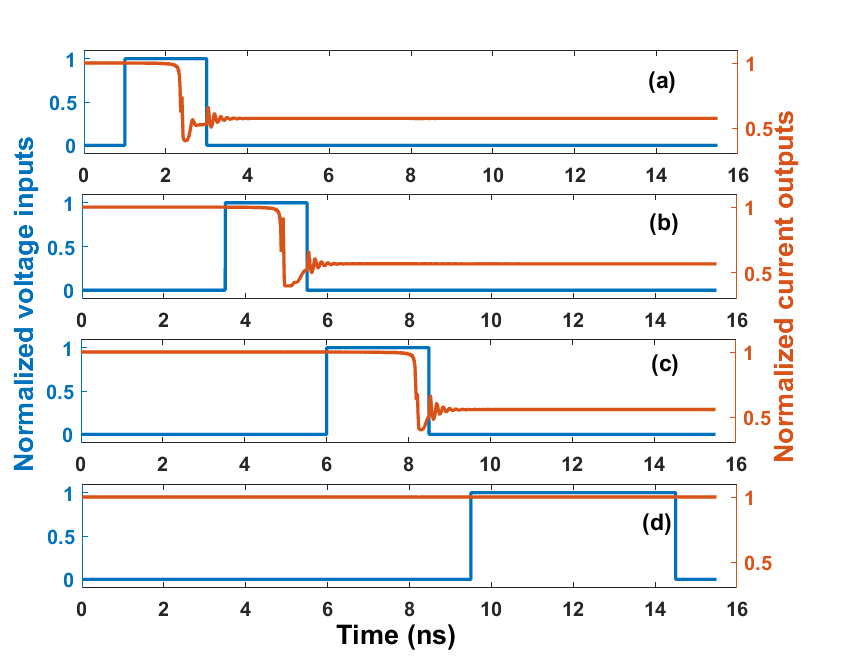}
\caption{Simulated electrical response at $\bm{A}$, $\bm{B}$, $\bm{{O}_\text{1}}$ and $\bm{{O}_\text{2}}$ of the bit-cell during write and $\it{XOR}$ operations. The output of $\it{XOR}$ computation between $\bm{A}$ = ‘1’ and $\bm{B}$ = ‘1’ is sensed from the current output at $\bm{{O}_\text{2}}$. Low normalized current outputs in $\bm{A}$, $\bm{B}$ and $\bm{{O}_\text{1}}$ corresponds to bit `1', whereas the high normalized current output in $\bm{{O}_\text{2}}$ corresponds to bit `0'.}
\label{fig:IVchar}
\addtocounter{figure}{0}
\end{figure} 

\begin{figure}[t]
\centering
\includegraphics[width=3in]{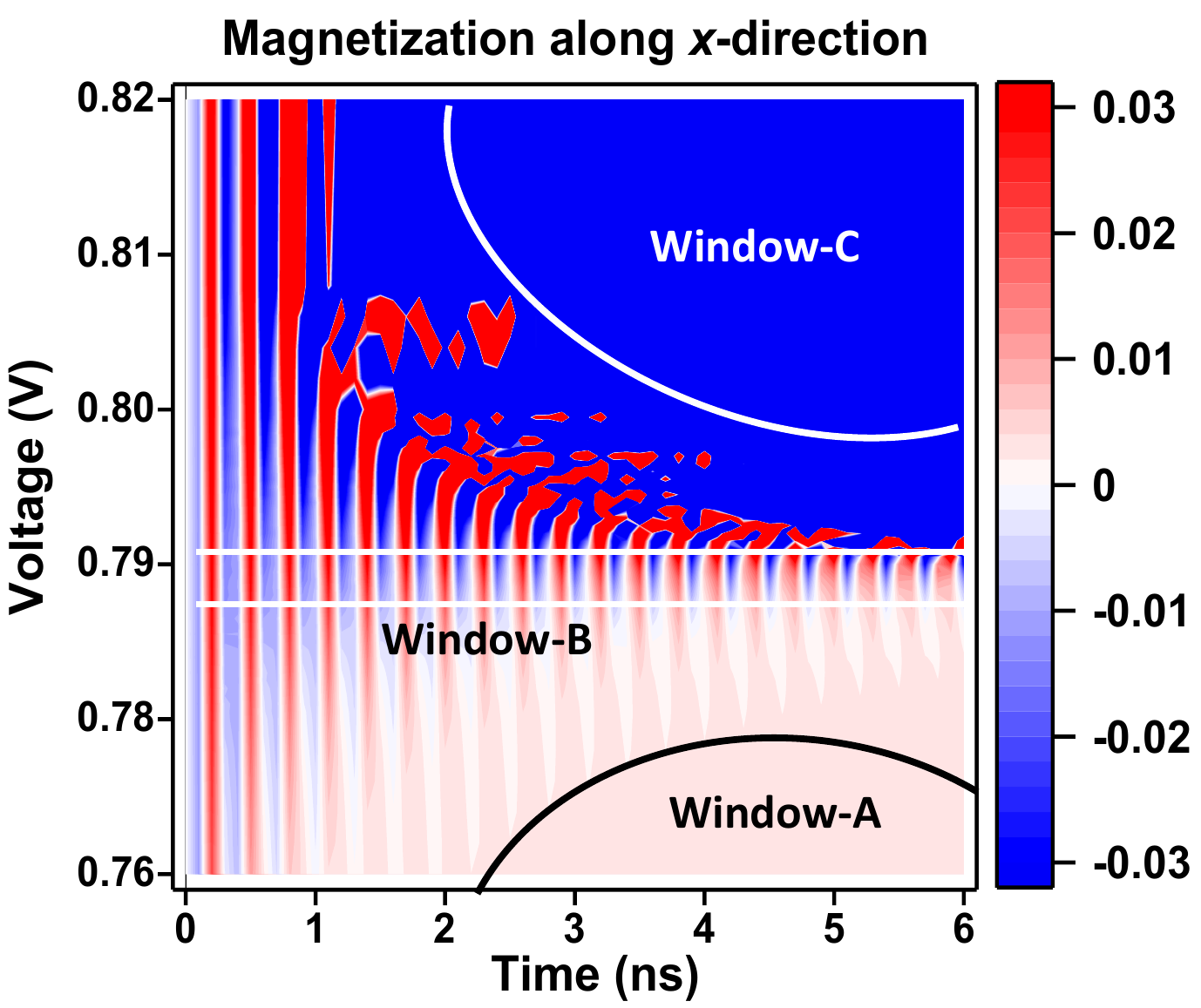}
\caption{The dependence of magnetization's $\it{x}$-component on input voltages with different pulse widths. The magnetic attenuation (window-A), stable oscillations (window-B) and skyrmion formation (window-C) regimes in the phase plot are labelled accordingly. Negative magnetization $\it{x}$-component in window-C denotes the nucleation of a skyrmion.}
\label{fig:phase}
\addtocounter{figure}{0}
\end{figure}

\subsection{Electrical Behavior of the Bit-cell}
\label{subsec:electbeh}

Fig.~\ref{fig:IVchar} shows the electrical behavior of the bit-cell during write followed by the $\it{XOR}$ operation. The electrical stimuli listed in ``Write ($\bm{A}$)'' row of Table~\ref{tab:inputs} is applied to the bit-cell when bit `1' is being written into $\bm{A}$. Consequently, a skyrmion is nucleated in $\bm{A}$ and ${R}_\text{A}$ increases, which reduces the magnitude of current flow in $\bm{A}$ (see Fig.~\ref{fig:IVchar}~(a)). The total delay for the write operation is about 2.5~ns (inclusive of 0.5~ns relaxation delay). Similarly, when bit `1' is being written into $\bm{B}$, the electrical stimuli listed in ``Write ($\bm{B}$)'' row of Table~\ref{tab:inputs} is applied to the bit-cell. Consequently, the skyrmion gets nucleated in $\bm{B}$ and the magnitude of current flow in $\bm{B}$ reduces (see Fig.~\ref{fig:IVchar}~(b)). The total delay of this write operation is also about 2.5~ns (inclusive of 0.5~ns relaxation delay).

The two-cycle $\it{XOR}$ operation is next performed by first activating the STNO at $\bm{{O}_\text{1}}$ to compute $\bm{A}\cdot\bm{B}$ and then activating the STNO at $\bm{{O}_\text{2}}$ in the second phase. For activating the STNO at $\bm{{O}_\text{1}}$, the electrical stimuli listed in ``Compute ($\bm{{O}_\text{1}}$)'' row of Table~\ref{tab:inputs} is applied to the bit-cell. As shown in Fig.~\ref{fig:IVchar}, when both the inputs, $\bm{A}$ and $\bm{B}$, are in state ‘1’, a skyrmion is nucleated in $\bm{{O}_\text{1}}$. As a result, ${R}_\text{O1}$ increases and current flow through $\bm{{O}_\text{1}}$ is reduced (See Fig.~\ref{fig:IVchar}~(c)). The total delay of this operation is about 3.0~ns (inclusive of 0.5~ns relaxation delay). In the final compute cycle, the STNO in $\bm{{O}_\text{2}}$ is activated for 5~ns using the electrical stimuli listed in ``Compute ($\bm{{O}_\text{1}}$)'' row of Table~\ref{tab:inputs}. Despite the voltage stimulus, no skyrmion gets nucleated in $\bm{{O}_\text{2}}$ due to the presence of a skyrmion in $\bm{{O}_\text{1}}$. Hence, there is no change in ${R}_\text{O2}$ and the magnitude of current flow through $\bm{O}_\text{2}$ (see Fig.~\ref{fig:IVchar}~(d)). In this case, the output at $\bm{O}_\text{2}$ is `0', which corresponds to the result of the $\it{XOR}$ operation between data stored at $\bm{A}$ and $\bm{B}$. After the compute operation has finished, the output at $\bm{O}_\text{2}$ can be read by applying the electrical stimuli listed in ``Read'' row of Table~\ref{tab:inputs} to the bit-cell.

The voltages applied to $\it{CSL}_\text{2}$ and $\it{CSL}_\text{3}$ during write and compute operations, respectively, are determined from the magnetization phase diagram shown in Fig.~\ref{fig:phase}, which was obtained by varying the voltage magnitude and pulse width. When the input voltage is less than 0.78~V, the magnetic oscillations attenuate within the time period of 4 ns (Window-A). If the input voltage is in between 0.788~V and 0.79~V, magnetization oscillates continuously for 6 ns (Window-B). Finally, if the voltage is greater than 0.8~V, magnetic oscillations are amplified and a skyrmion is nucleated (Window-C). Therefore, we apply 0.81~V to ${CSL}_\text{2}$ during write operation, and 0.79~V to ${CSL}_\text{3}$ during compute operation (Table~\ref{tab:inputs}). The voltage used for the read operation (\emph{i.e.}, 0.25~V) is selected to avoid read-failure and read-disturb errors.

\begin{table}[t]
\centering
\caption{Performance Evaluation of 1~KB SIMC}
\includegraphics[width=2.5in, clip, trim=0.25in 0.17in 0.2in 0.17in, scale=1.0]{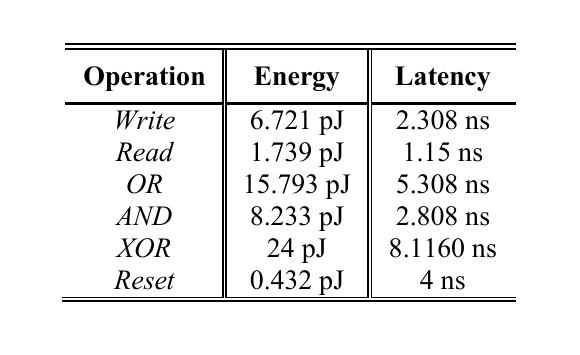}
\label{tab:SIMC}
\end{table}

\subsection{Evaluation of SIMBA Hardware Architecture}
\label{subsec:layout}

Based on the device write current requirements and FreePDK45 design rules, the area of each bit-cell in the SIMC is about 900~F$^{2}$ (Fig.~\ref{fig:layout}~(a)). We note that the bit-cell area is limited by the transistor size due to the high write current requirements ($\approx$~490~$\mu$A) of the memory device. Using the extracted electrical characteristics of an individual bit-cell, an SIMC with 1~KB data storage capability was simulated in NVSim to estimate the energy consumption and latency for different operations, which are summarized in Table~\ref{tab:SIMC}. The simulation results for the 1~KB SIMC units are used to evaluate the overall SIMBA hardware architecture. 

In addition, the electrical stimuli applied to the $\it{popcount}$ and $\it{comparator}$ circuits proposed in Fig.~\ref{fig:SIMBA}~(c) are listed in Table.~\ref{tab:pinputs}. Since the switching characteristics of SOT-MTJs are well known in literature \cite{Garello2018, Xin2018, Parveen2017}, the simulation results of $\it{popcount}$ and $\it{comparator}$ circuits have been omitted. The energy and latency consumed during these operations are determined to be 7.5~pJ and 10~ns, respectively. A $\it{popcount}$ circuit processing 100 bits at a time is used for estimating these energy and latency values.

\begin{table}[t]
\centering
\caption{VGG-like BNN Model}
\includegraphics[width=2.75in, clip, trim=0.25in 0.17in 0.2in 0.17in, scale=1.0]{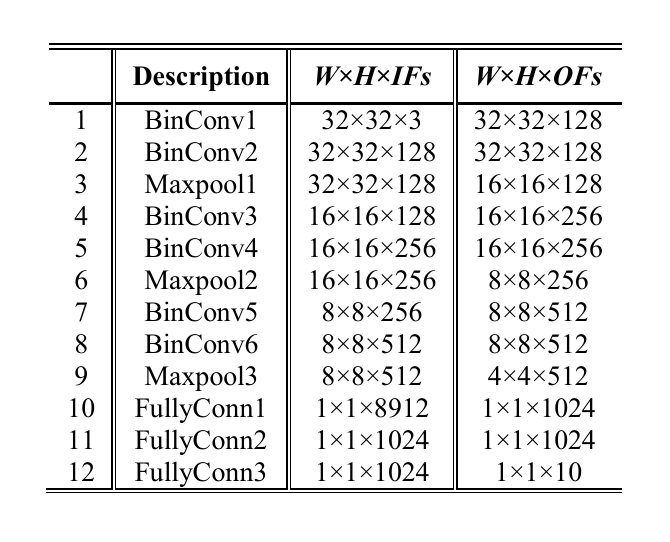}
\label{tab:layers}
\end{table}

\begin{figure}[t]
\centering
\includegraphics[width=3.5in]{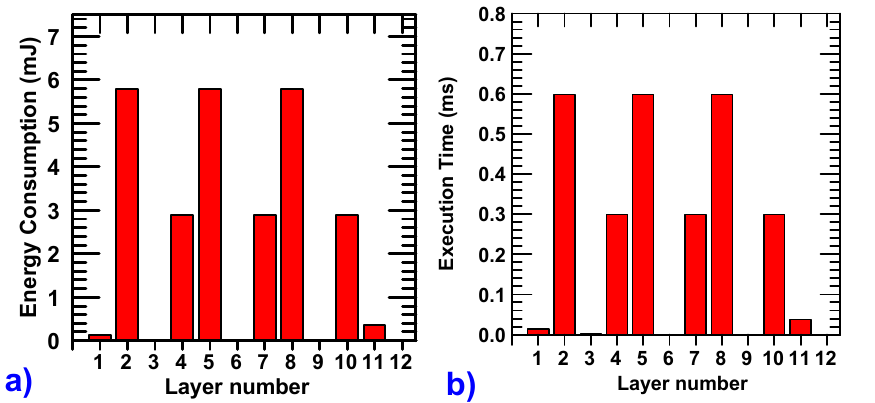}
\caption{Estimated layer-wise (a) energy consumption and (b) execution times per inference in SIMBA.}
\label{fig:layer}
\addtocounter{figure}{0}
\end{figure}

\begin{figure*}[t]
\centering
\includegraphics[width=5in]{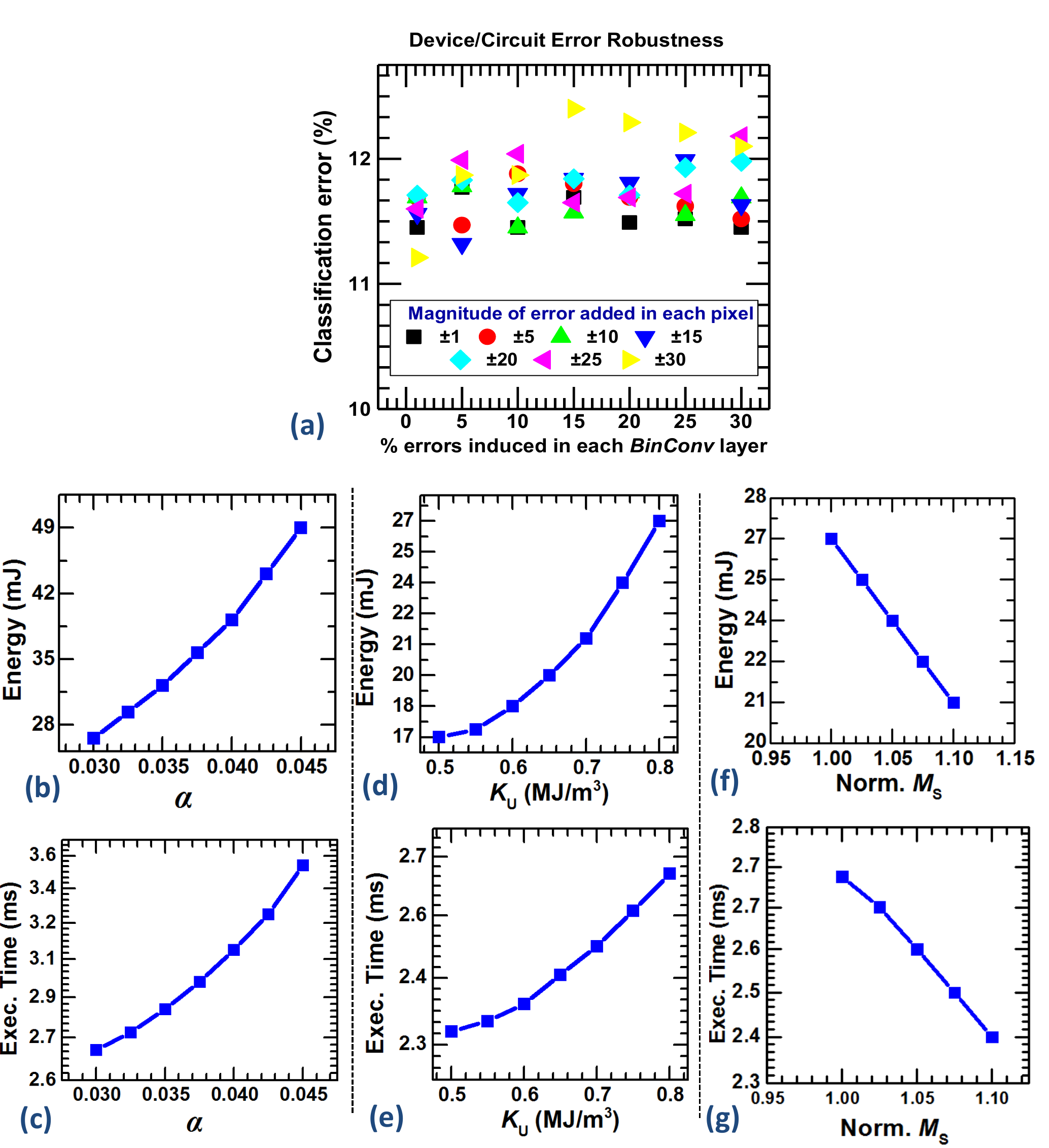}
\caption{(a) Variations in error rate per inference induced by the stochastic behavior of devices/circuits used in the hardware. (b-g) The influence of material properties on the performance of SIMBA (energy consumption and execution time). (b-c) damping ratio, $\alpha$, (d-e) anisotropic energy density, $K_\text{U}$, and (f-g) saturation magnetization, $M_\text{S}$.}
\label{fig:param}
\addtocounter{figure}{0}
\end{figure*}

\begin{table*}[!t]
\centering
\caption{Electrical inputs for the $\it{popcount}$ and $\it{comparator}$ circuits}
\includegraphics[width=5in, clip, scale=1.0]{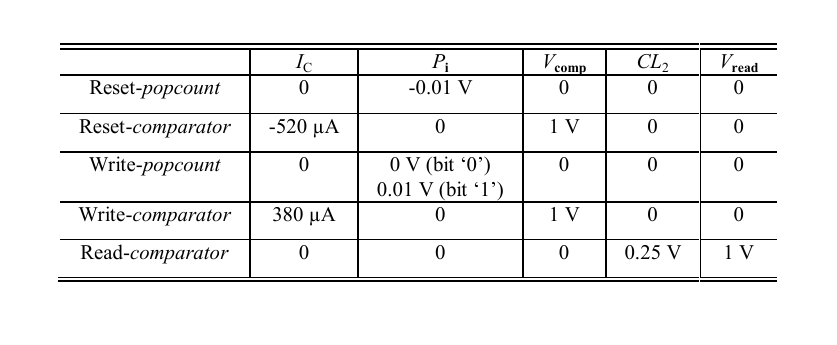}
\label{tab:pinputs}
\end{table*}

In this work, we designed SIMBA to accelerate the computations for a VGG-like BNN with 12 layers (see Table~\ref{tab:layers}) \cite{Cour2016, Guo2018}. The VGG-like BNN model performs the classification task on CIFAR-10 image data with an accuracy of 88.5\%. Each BinConv layer in the network performs i) $\it{kw~\times~kh~\times~OF~\times~W~\times~H~\times~IF}$ number of $\it{XOR}$ operations, ii) $\it{W~\times~H~\times~OF}$ number of $\it{popcount}$ operations (each on $\it{kw~\times~kh~\times~IF}$ number of bits), and iii) $\it{W~\times~H~\times~OF}$ number of $\it{comparator}$ operations. Each Maxpool layer performs $\text{3}\it{~\times~W~\times~H~\times~OF}$ number of 2-bit $\it{OR}$ operations, where the maxpooling filter size is 2$\times$2. Each FullyConn layer performs i) $\it{IF~\times~OF}$ number of $\it{AND}$ operations, ii) $\it{popcount}$ operation on $\it{IF}$ number of bits, and iii) $\it{OF}$ number of $\it{comparator}$ operations. Using the results in Table~\ref{tab:SIMC} and results from circuit and device simulations, the estimated energy consumption and time delay in each BNN layer are shown in Fig.~\ref{fig:layer}, respectively. Over all, SIMBA is estimated to require 26.7~mJ of energy and 2.7~ms time delay to perform one classification task, which translates to throughput of 370.4 (images/sec). 

\subsubsection{Analysis of Device Characteristics on SIMBA Performance}

We have also explored the impact of material properties (such as damping ratio, $\alpha$, anisotropic energy density, $K_\text{U}$, and saturation magnetization, $M_\text{S}$) on the performance of proposed hardware. We found that a 33.3\% of reduction in $\alpha$ improves the energy efficiency by 47\% with 1.3$\times$ speedup (Fig.~\ref{fig:param}~(b-c)). Reduction in $\alpha$ decreases the current required to drive magnetization oscillations and improves energy efficiency and execution times. Separately, over 37.5\% reduction of $K_\text{U}$ improves the energy efficiency by 36.84\% with 1.18$\times$ speedup (Fig.~\ref{fig:param}~(d-e)). Similar to the reduction in $\alpha$, the current required to induce magnetization oscillations or switching reduces with reduced $K_\text{U}$. When only $M_\text{S}$ is increased by 10\%, the energy consumption is reduced by 23.68\% with speedup of 1.13$\times$ (Fig.~\ref{fig:param}~(f-g)). This is due to increase in the demagnetizing field as $M_\text{S}$ increases. The demagnetizing field tends to reduce 
the current needed to induce magnetization oscillations or switching.

Since thermal effects can affect the magnetization dynamics of the spintronic devices and introduce errors in SIMBA, there is a need to study the influence of hardware errors on the overall classification accuracy. We induced errors in 1\%-30\% of the pixels in each BNN layer and observed the overall classification accuracy. Note that the magnitude of error in each pixel is randomly varied from $\pm\text{1}$ to $\pm\text{30}$. As shown in Fig.~\ref{fig:param}~(a), the classification accuracy degrades by only 1\% despite the presence of 30\% errors in each BinConv layer. The worst classification error rate we observed is 12.4\% whereas the error rate without thermal effects is around 11.5\%.

%% file: sections/VI/Sec_VI.tex
\section{Conclusion}
\label{sec:conclusion}

In summary, we propose a binary neural network (BNN) accelerator, referred to as SIMBA, based on a novel non-volatile skyrmionic in-memory processing engine. A hybrid mixed-mode device-to-architecture simulation framework was developed to design and evaluate SIMBA. Simulation results show that SIMBA consumes 26.7~mJ of energy and 2.7~ms of delay per inference task on a VGG-like BNN. The inference error rate is less than 1\% even in the presence of 30\% bit errors in the skyrmionic devices. Finally, we showed improvements in the performance of SIMBA that can be achieved by tuning device material parameters such as damping ratio, anisotropic energy, and saturation magnetization.